Evaluation of the Influence of Structural Parameters on the Mechanical Properties of Foam Glasses via *In-Situ* Micro-CT Mechanical Testing

Mateus Gruener Lima[1,2], Tobias Günther[3], Thu Trang Võ[4], Eduardo Inocente Jussiani[1], Dirk Enke[3], Urs A. Peuker[4], Ralf B. Wehrspohn[5], Juliana Martins de Souza e Silva[2,5,*]

[1] Applied Nuclear Physics Research Group, State University of Londrina, Londrina, Brazil
[2] Fraunhofer Institute for Microstructure of Materials and Systems IMWS, Halle (Saale), Germany
[3] Institute of Chemical Technology, Leipzig University, Leipzig, Germany
[4] Institute of Mechanical Process Engineering and Mineral Processing, TU Bergakademie Freiberg, Freiberg, Germany
[5] Institute of Physics, Martin Luther University Halle-Wittenberg, Halle (Saale), Germany
* juliana.martins@physik.uni-halle.de

**Abstract**

Foam glass made from waste glass has high chemical and mechanical stability and flexible structural properties, making it suitable for a wide range of applications. To ensure its reliability, it is essential to understand its mechanical properties and fracture mechanisms. In this study, we investigated morphological features related to pore structure alongside mechanical properties, specifically Young's modulus and compressive strength, of three different monolithic foam glass samples using micro-computed tomography combined with *ex-situ* and *in-situ* uniaxial compression experiments and Digital Volume Correlation (DVC) analysis. The foam glasses exhibit an inverse relationship between mechanical strength and factors such as wall thickness, porosity, pore size and irregularity, with the compressive strength following a power-law correlation with the proportion of large pores. The multi-peak behavior of the stress-strain curves indicates micro-cracking within the porous lattice. Micro-CT data show that damage is concentrated at the upper and lower extremes of the specimens, and that the applied strain induces changes in the porosity, mean pore diameter, and pore sphericity, driven by deformation and collapse of the pore structures. DVC analysis quantitatively validated these observations.

# 1. Introduction

Materials with porous lattices are widely recognized for their versatile properties and applications in fields such as civil engineering [1], thermal insulation [2,3], biomimetics [4–6], and chemical separation [7–9], including $CO_2$ filtration [10]. Among these, foam glasses synthesized by a one-step foaming process are notable for their light weight character, high mechanical and chemical stability, and excellent thermal insulation properties [11,12]. Their main advantage is that they can be produced partly or exclusively from different types of waste glass [13–15].

To produce foam glass, fine glass powder is mixed with a suitable foaming agent, a compound that releases gas at high temperatures. When the powder mixture is heated, the glass particles sinter together while the foaming agent releases gas, leading to the formation of a pore system [15]. Although the optimal foaming agent depends on the glass composition, commonly used options include carbonates (namely $Na_2CO_3$ [16,17] and $CaCO_3$ [18,19]), sodium silicate water glass [20,21] and various manganese oxides ($Mn_2O_3$, $MnO_2$, $Mn_3O_4$) [22]. Due to the potential application of foam glass as an insulating material, research is focused on minimizing thermal conductivity and apparent density and increasing mechanical stability. König *et al.* [23] investigated the thermal conductivity of foam glasses made from lead-free CRT panel glass, $MnO_2$ and C, finding values between 41 and 60 mW m$^{-1}$ K$^{-1}$ at densities ranging from 115 and 200 kg m$^{-3}$. Notably, higher densities were linked to increased thermal conductivities. Tulyaganov *et al.* [24] assessed the compressive strength of foam glasses made from soda-lime-silica-glass and SiC, among other components, reporting values up to 2.6 MPa at an apparent foam glass density of 252 kg m$^{-3}$. König *et al.* [25] evaluated the mechanical strength of foam glasses made from waste soda-lime-silica-glass, $Mn_3O_4$ and C, noting strengths around 0.6 MPa for apparent densities near 120 kg m$^{-3}$. Another study of commercial low-density foam glass (115 kg m$^{-3}$) revealed a compressive strength of approximately 0.4 MPa [26]. Commercial foam glasses from GLAPOR [14] made from waste car window glass, glycerol, and possibly other undisclosed foaming agents, have thermal conductivities of 52 to 58 mW m$^{-1}$ K$^{-1}$, and compressive strengths of 0.6 to 1.6 MPa with densities of 130 to 155 kg m$^{-3}$.

The mechanical stability of foam glass is essential to ensure resilience, load-bearing capacity and fatigue resistance, but also to enable design optimization and appropriate material selection. Their performance is strongly dependent on the stiffness of the solid phase and the morphology of the pore network. To understand the morphology and mechanical properties of porous materials in detail, imaging techniques such as X-ray micro-computed tomography (micro-CT) with *in-situ* mechanical testing [27–29], and analytical methods such as Digital Volume Correlation (DVC) analysis [30–32] are used. These approaches provide access to quantitative structural parameters, including porosity, pore size distribution, and the thickness of internal walls within the pore lattice. In addition, DVC allows real-time monitoring of material's deformation during mechanical testing, enabling the evaluation of mechanical properties, such as displacement and strain of internal features under applied loads [33,34], and providing immediate feedback on how morphological changes affect the material performance. The combination of these experimental and analytical techniques provides fundamental insight into the mechanical stability and fracture behavior of porous materials, aids design changes that affect material performance, and allows rapid iteration and optimization for appropriate use in a wide range of applications. Therefore, in this study we aimed to comprehensively investigate the structural and mechanical behavior of three foam glass samples with similar chemistry and porosity, but with different morphological properties, namely pore diameter and porous wall thickness. By integrating mechanical testing with micro-CT, and by applying DVC analysis to the *in-situ* mechanical test data obtained, we sought to elucidate the relationships between morphological characteristics and mechanical performance, in particular Young's modulus and compressive strength. This approach not only improves our understanding of the foam glass but also provides a framework for optimizing its design and application in various fields.

## 2. Experimental methods

### 2.1 Porous glass monoliths synthesis

Open porous foam glasses were synthesized by a modified route from König *et al.* [25]. Waste flat glass from car windows (supplied by Franz Rottner bi-foam Schaumglas GmbH, Oranienbaum-Wörlitz, Germany) with a particle size ($d_{90}$ = 95 µm) was mixed with 6.00 wt. % $Mn_2O_3$ (98 %, Thermo Scientific), 0.28 wt. % $TiO_2$ (99 %, Merck) and homogenized in a ball mill (PM 100, Retsch) at 450 rpm for 10 min. Then, 45 g of the obtained powder was thoroughly mixed with 1.00 - 1.60 g glycerol (86 - 88 %, Acros Organics) and 5.50 g of fully desalted water before being transferred to 8 × 8 cm stainless steel molds, covered with mineral fiber fleece. Molded powders were heated in a muffle furnace (B180, Nabertherm) under air atmosphere at 20 °C/min to 810 °C and foamed at this temperature for 20 - 30 min. The samples were allowed to cool to room temperature before further testing. The exact synthesis conditions for each sample are listed in Table S1.

From each porous foam glass sample (referred here as PG_A, PG_B and PG_C), three specimens were cut into cuboids with rough dimensions of 8 mm × 5 mm × 5 mm and then sanded to 7 mm × 4 mm × 4 mm. Two of those specimens were allocated for *ex-situ* uniaxial compression, while the third was reserved for *in-situ* uniaxial compression in a micro-CT scan.

### 2.2 *Ex-situ* mechanical testing

*Ex-situ* compression of the monoliths (Table S2) was carried out in duplicate using a Texture Analyser (TA.XTplus, Stable Micro Systems). A constant compression rate of 1.2 mm min$^{-1}$ was applied to the longer axis, and testing continued until a strain of 20 % was reached for all samples. The Young's modulus $E$ for each specimen was determined as the slope of a linear, stable section of the stress-strain curves. The maximum compressive strengths $\sigma_{max}$ and the mean Young's modulus for each specimen were calculated (Table S3) with standard deviation between the two measurements.

## 2.3 *In-situ* intermittent mechanical testing in a micro-CT

A CT5000 *in-situ* load cell (Deben UK Ltd) was used to mechanically compress the foam glass, with the lower piston moving upwards to compress the sample (y-axis). It consists of two flat ceramic pistons (xz) between which the sample is positioned. A minimum force of 2 N was applied to the sample to ensure stable positioning prior to the start of the compression experiment, and the load cell was positioned in the direction (z-axis) of the X-ray beam of a Carl Zeiss Xradia 510 Versa. The samples were scanned with an applied voltage of 80 kV, 7 W of power, no radiation filter, and an additional optical magnification of 0.4 ×. Each scan consisted of 801 projections of 12 s each, for a total scanning time of approximately 2:40 h and an image pixel size of 10 μm. Five scans were performed for each specimen: a reference scan with a 2 N force applied to lightly touch and hold the specimen in position. This scan was followed by four intermittent subsequent scans at incremental load steps defined at strain values of 0.5 %, 1.0 %, 2.0 %, and 4.0 % calculated relative to the specimen height, at a constant compression rate of 1 μm/s. All datasets were reconstructed using the XMReconstructor software (version 11.1.8043). Identical reconstruction parameters were used for all measurements. The correction factor for the beam hardening artefact was 0.05, and the byte scaling ranged from -0.1 to 0.25.

Images from the reference scan were used for structural characterization of the samples, focusing on the pore lattice and foam structure (Fig. S1 and Table S4). Excluding pores with diameters smaller than 30 μm (3 × pixel size [35,36]), porosity was determined as the ratio of the total pore volume to the total volume of interest considered in the analysis. After characterizing each sample, the same image processing procedure was applied to the following four datasets for each sample to assess the effects of uniaxial compression on the foam structure and to enable DVC analyses to evaluate displacement fields and strain maps corresponding to each load step (Fig. S2).

## 3. Results and discussion

In this work, we have analyzed three porous glass samples (PG_A, PG_B, and PG_C), which have the same chemical composition but different morphological properties. We used micro-CT for their detailed morphological characterization, *ex-situ* and intermittent *in-situ* mechanical testing in a micro-CT for their mechanical characterization (Fig. 1). Although the samples have similar visual appearance (Fig. 1 a), a detailed structural characterization using micro-CT, image processing and morphological analysis (Fig. 1 b) shows that the samples have different morphological characteristics defined by different porosity and pore size distribution (Fig. 2). Sample PG_A has the largest pores, with diameters ranging from 0.03 to 1.09 mm. Sample PG_B has pore diameters ranging from 0.03 to 1.00 mm, while sample PG_C has a narrower range from 0.03 to 0.83 mm (Fig. 2 d-f).

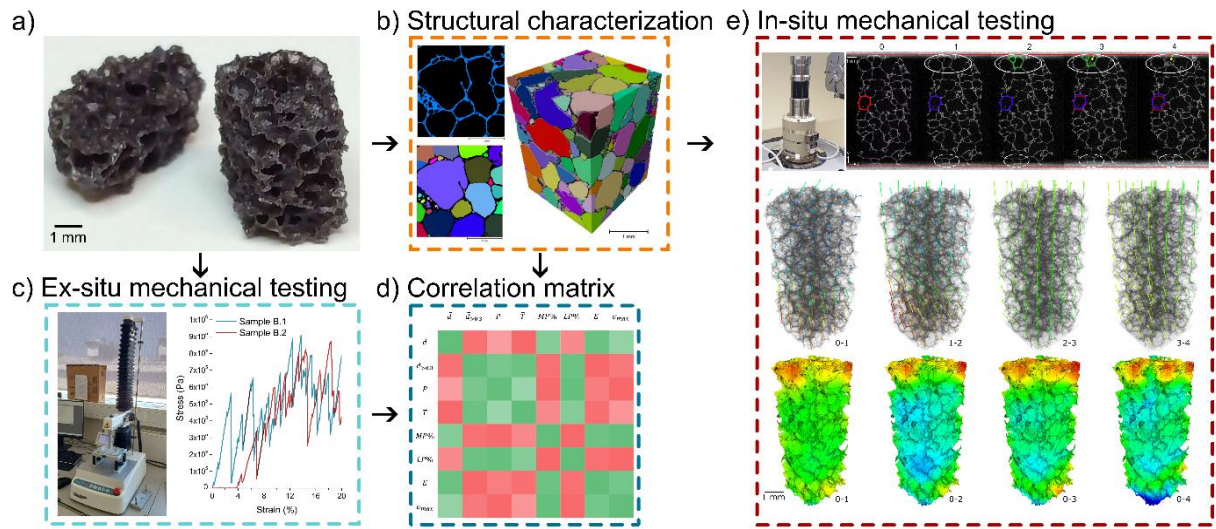

Fig. 1. Overview of this work's analysis workflow, showing a) the samples' appearance after cutting and sanding them for the mechanical testing performed. Prior to any mechanical stress, the samples were b) imaged using a micro-CT and structural characterization was performed. c) *Ex-situ* uniaxial compression was performed and d) the degree of correlation between structural parameters and mechanical behavior was stablished. *Ex-situ* mechanical data were used to guide the parameters for e) the *in-situ* uniaxial compression, which was performed in a micro-CT the volumetric images obtained were analyzed using DVC.

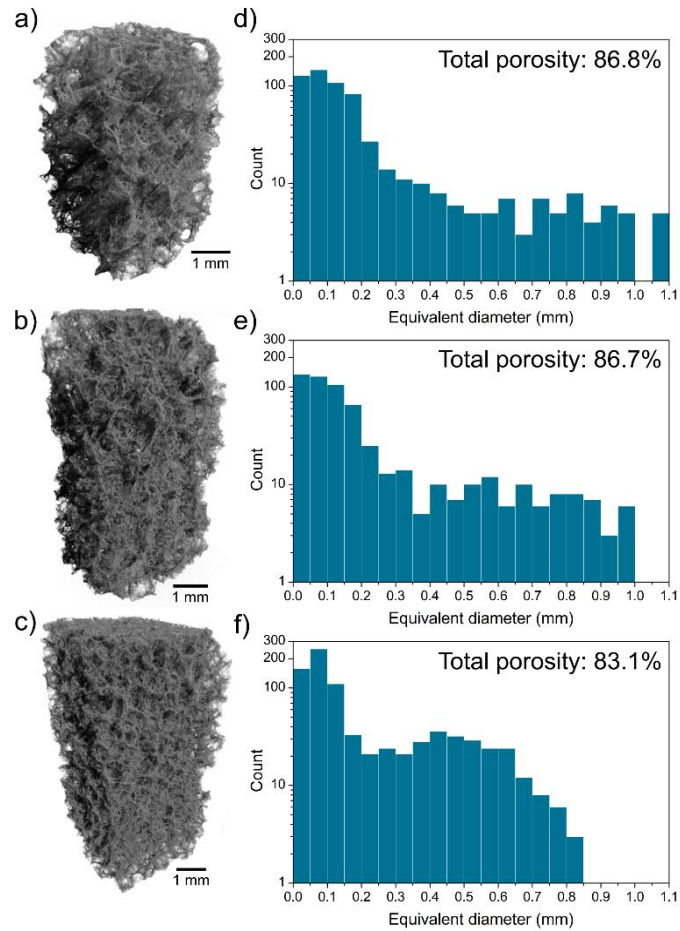

Fig. 2. Three-dimensional reconstruction of micro-CT scanned samples: a) PG_A, b) PG_B, c) PG_C, along with their respective pore equivalent diameter distributions in d), e), and f). Insert: total porosity value.

To assess their mechanical behavior, all porous glass samples were subjected to standard *ex-situ* mechanical compression tests (Fig.1 c, Fig. 3 a-c). All samples exhibited small sudden drops in the stress-strain profile, which are typically indicative of microcracking or localized failure events. Similar multi-peak stress-strain profiles have been reported in previous studies on various porous glass and ceramic scaffolds [37,38]. These observations suggest that, as compression progressed, multiple small fractures formed, redistributing forces among the remaining structures. This process eventually leads to more significant fractures, as seen in the large stress drop in the 11 % - 12 % strain range for specimen PG_C1 (Fig. 3 c).

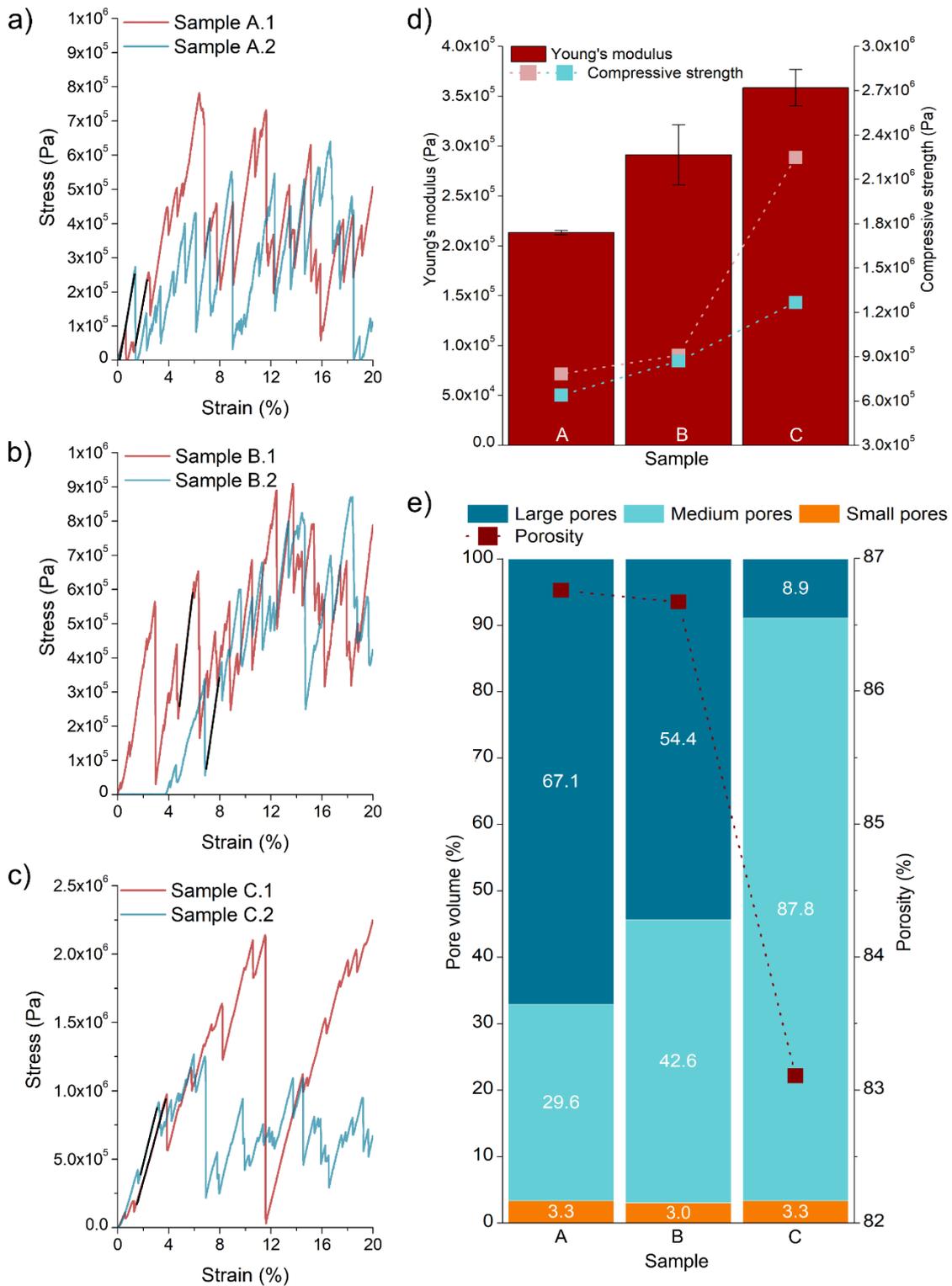

Fig. 3. Stress-strain curves obtained from *ex-situ* compression experiments (duplicates) on samples a) PG_A, b) PG_B, and c) PG_C. Black lines indicate the linear region used to calculate Young's moduli. d) Mean Young's modulus, with standard deviation as error bars, and maximum compressive strength (right axis) for two specimens of each porous glass sample. e) Pore volume distribution (left axis) and total porosity (right axis) of each sample, derived from micro-CT reference scan analysis. Pores are categorized by equivalent diameter: small (< 0.3 mm), medium (0.3 - 0.8 mm), and large (> 0.8 mm).

Samples PG_A and PG_B exhibit similar porosities, while PG_C has the lowest porosity among the samples. The mean Young's modulus $E$ and maximum compressive strength $\sigma_{max}$ of the porous glass samples exhibit an increasing trend from PG_A to PG_B and PG_C (Fig. 3 d), with statistically significant differences. The $E$ values, ranging between 0.2 and 0.4 MPa, and $\sigma_{max}$ values, approximately in the range of units of MPa, align with those reported in previous studies on foams with 80 - 90 % porosity [39–42].

Although the porosities of the three samples range between 83 % and 87 %, the distribution of pore sizes varies significantly. When separating the pores into small (equivalent diameter less than 0.3 mm), medium (0.3 to 0.8 mm), and large (diameters greater than 0.8 mm) categories, notable differences are observed in the distribution of medium and large pores among the three samples (Fig. 3 e). While all three samples exhibit similar percentages of small pores, the proportion of medium pores increases as the percentage of large pores decreases in PG_A, PG_B, and PG_C, respectively.

Considering the predictions of a power-law dependency of Young's modulus or compressive strength on porosity [43–46], we compared the dependency of the mean compressive strength $\bar{\sigma}$ to the large pores' porosity of the samples, using the following relation:

$$\sigma = C \left( -\frac{aP_V}{100} \right)^n \tag{1}$$

where $\sigma$ is the compressive strength of the foam glass, $P_V$ is the large pores' porosity, $C$, $a$ and $n$ are constants and can be extracted from the slope of a *semi-log* plot of $\sigma$ to $P_V$. In this study, the porous foam glass samples showed a similar trend to that reported in [45,46], but only when considering the porosity of the larger pores (Fig. 4). Thus, an increase in total porosity, particularly the porosity of the large pores, correlates with a reduction in mechanical strength with an exponential decrease in compressive strength. This observation aligns with findings from previous studies for other glass [42], ceramic [41,47], and metallic scaffolds [48–50].

The compressive strength of our samples is also influenced by the distribution of pore sizes (Fig. 2 d - f). Research has shown that a more uniform pore distribution is associated with greater resistance in foam glasses [51]. In our case, the pores in PG_C are distributed within a narrower size range, indicating a higher level of uniformity. This uniform pore distribution is likely a key contributor to PG_C's enhanced compressive strength.

Despite previous studies indicating that an increase in the thickness of internal walls enhances mechanical resistance [52], our experiment revealed an inverse relationship between mean wall thickness and the resistance parameters, $E$ and $\sigma_{max}$ (Table S5). This phenomenon can be attributed to a higher likelihood of micro-cracking due to increased bending stresses, which accumulate and lead to more extensive damage, as reported in studies on porous ceramics [53,54].

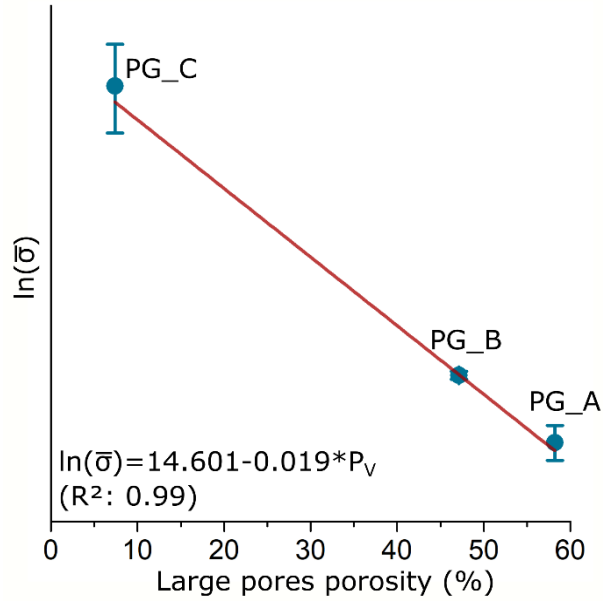

Fig. 4. Semi-log plot of the mean compressive strength $\bar{\sigma}$ versus the porosity of large pores $P_V$, fitted with a linear regression. The inset formula shows the regression equation and the corresponding R² value.

The *in-situ* micro-CT experiments allow visualization of changes in pore shape and position due to compression, as shown in the virtual slices across the five loading steps (Fig. 2 d, 0 - 4 in Fig. 5). Virtual slices illustrate the upward shift of pores (Fig. 5, blue highlight in 1 - 4 as an example) relative to the uncompressed specimen (Fig. 5, red highlight in 0 as an example), consistent with the experimental setup where the lower piston moves upwards to compress the sample. Indeed, the comparison of the two-dimensional (xz) volume fraction reveals more pronounced differences at the sample extremities, particularly within the first 100 slices (Fig. S3). For all samples, wall fractures occurred between successive load steps (Fig. 5, white ellipses). Fractures predominantly occur at the extremities of the samples, particularly at the contact points of the sample with the top and bottom pistons used for compression. They appear to develop gradually with each compression step, aligning with the micro-cracking behavior observed in the stress-strain curves from the *ex-situ* experiments (Fig. 3). The progression of local damage culminating in larger fractures aligns with observations reported in studies on ceramic and foam glasses [28,55]. Additionally, these slices reveal that the largest damages occur in sample PG_A, followed by sample PG_B and PG_C. This pattern is consistent with the values of $E$ and $\sigma_{max}$ determined earlier in the *ex-situ* experiments (Fig. 3 d), as these two quantities are indicative of higher mechanical stability. As expected, the samples with lower elastic modulus ($E$) and maximum stress ($\sigma_{max}$) - namely PG_A and PG_B - exhibited the most significant damage (indicated by solid white ellipses in Fig. 5). Upon closer inspection, we observed that broken fragments often filled the pores (highlighted by yellow arrows in Fig. 5) or came to rest on the bottom piston (steps 3 and 4, Fig. 5 a).

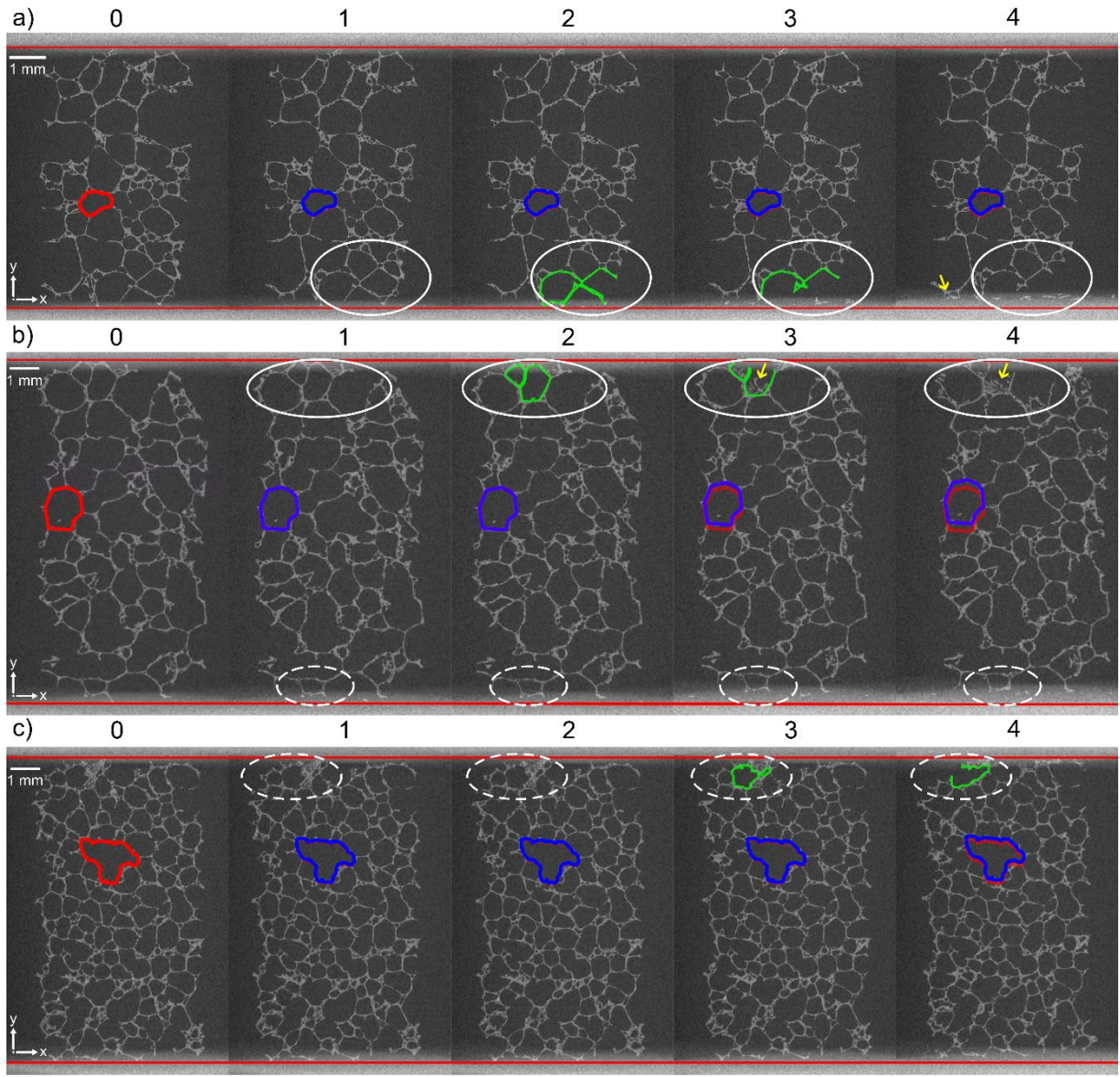

Fig. 5. Virtual slices of samples a) PG_A, b) PG_B, and c) PG_C obtained from the *in-situ* uniaxial compression experiments performed in the micro-CT. The slices represent the reference uncompressed position (0) and the four subsequent load steps (1-4). The red highlights indicate a reference pore in the initial dataset of the uncompressed samples, while the blue highlight indicates the same pore modified by compression. White ellipses denote fracture regions between load steps, with dashed ellipses marking small fractures and solid ellipses indicating larger breakage. Green highlights the pores that merged due the collapse of internal walls. Yellow arrows point to broken fragments that fell into the porous structure.

The effect of strain on each sample is indicated by changes in porosity, mean equivalent diameter and mean sphericity of the pores with increased strain (Fig. 6). For samples PG_A and PG_C, a decrease in porosity is observed after compression at low strains (Fig. 6 a). However, during subsequent compressions, including for sample PG_B, the porosity increases. This initial decrease in porosity can be attributed to densification of the pores caused by the lower forces applied during the first compression, as observed within the dashed white ellipses throughout steps 0 - 3 in Fig. 5 c. As higher forces are applied in subsequent steps, fractures occur, leading to pore merging (as highlighted in green in Fig. 5), which increases both the porosity and the pore mean equivalent diameter (Fig. 6 b). Both the pore densification and pore merging/collapse phenomena have been previously reported by others [32,56]. Additionally, a general decrease in the pore sphericity values of the specimens under compression (Fig. 4 c) is consistent with the ongoing compression deforming the pores further, shifting them away from a spherical shape and resulting in a subsequent decrease in sphericity. Previous studies have also highlighted a similar deformation process that occurs in pore scaffolds of various materials during compression [56,57].

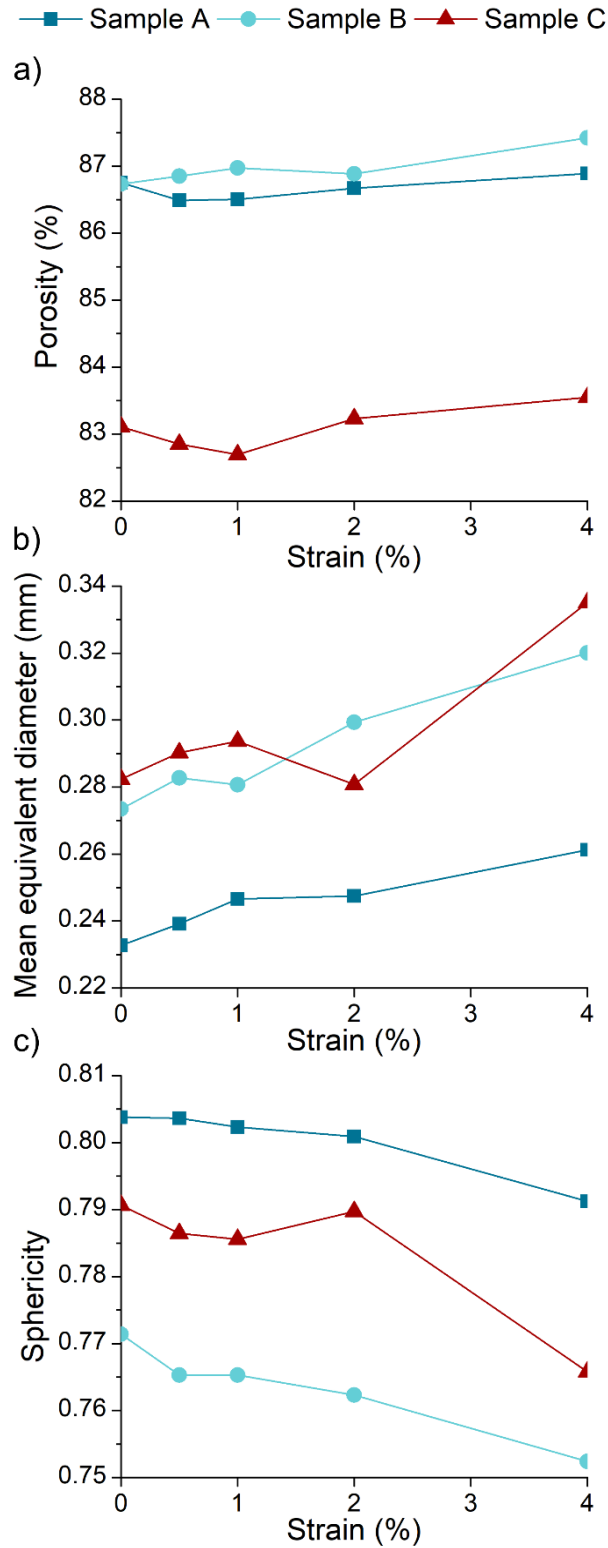

Fig. 6. Effect of increase of strain on samples' a) porosity, b) mean pore equivalent diameter, and c) mean sphericity of the pores.

We used DVC analysis to quantify morphological changes in the samples throughout the compression process during the *in-situ* experiment. Displacement fields were estimated for each load step (Figs. 7 a, 8 a, and 9 a), along with cumulative three-dimensional strain maps obtained after each compression step (Figs. 7 b, 8 b, and 9 b). Small displacement vectors are visible during the initial compressions (Figs. 7 a, 8 a, and 9 a, steps 0 - 1 and 1 - 2). In PG_B (Fig. 8), larger vectors are observed near the bottom region, probably due to sample rearrangement. As higher compressions are applied (steps 2 - 3 and 3 - 4 in Figs. 7 - 9), the displacement vectors become more evenly distributed with larger magnitudes, corresponding to 1 % and 2 % deformations of the total sample height, respectively [58].

The strain maps from the DVC analysis show that regions with high strain magnitudes (both positive and negative) are mainly located at the upper and lower extremes of the samples (Figs. 7 b, 8 b, and 9 b), as previously observed in the comparison of the reconstructed slices (Fig. 5), where these regions showed more damage. These high-strain areas are associated with larger fractures, in contrast to the central region, where the strain is more uniformly distributed, and minimal damage was observed. This pattern aligns with findings from previous studies, which have reported a relationship between high-strain regions and structural collapse in various porous materials [32,59,60].

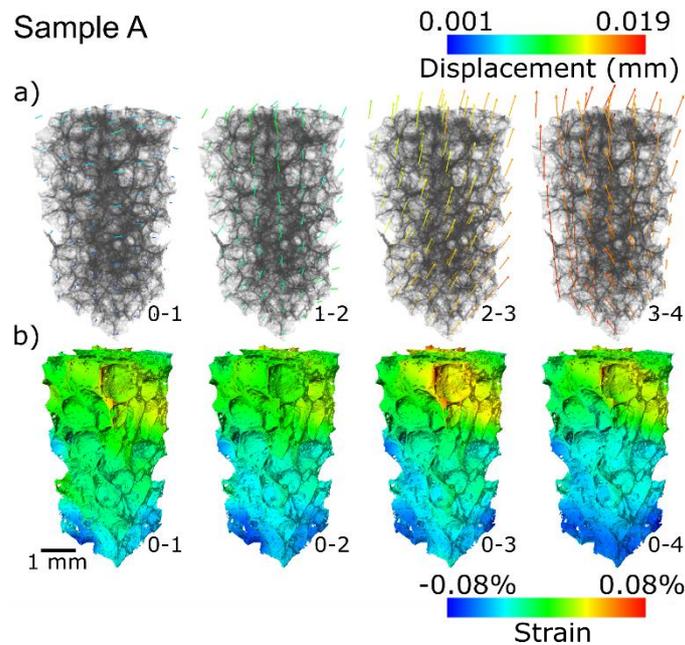

Fig. 7. a) Displacement vectors and b) strain maps for each load step in sample PG_A, analyzed using Digital Volume Correlation from the *in-situ* compression test micro-CT scan data.

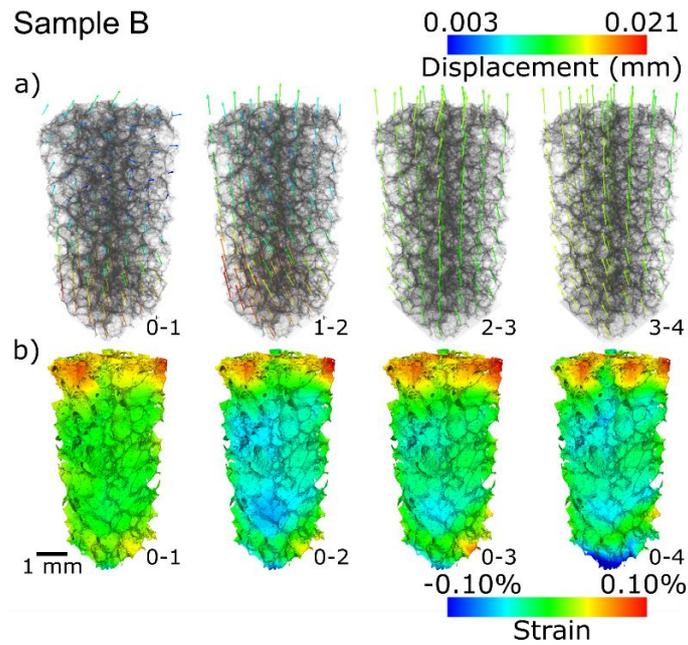

Fig. 8. a) Displacement vectors and b) strain maps for each load step in sample PG_B, analyzed using Digital Volume Correlation from the *in-situ* compression test micro-CT scan data.

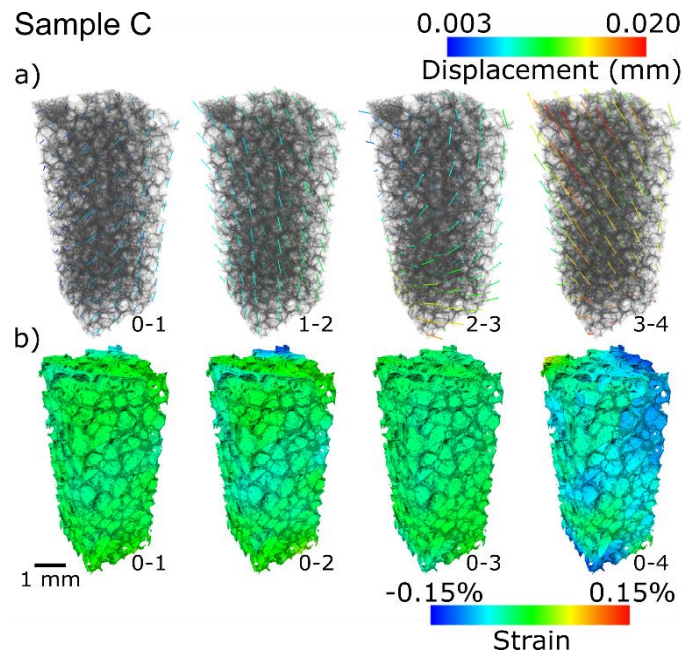

Fig. 9. a) Displacement vectors and b) strain maps for each load step in sample PG_C, analyzed using Digital Volume Correlation from the *in-situ* compression test micro-CT scan data.

## 4. Conclusion

The use of porous foam glasses in various fields requires an understanding of their mechanical properties and the identification of potential fracture conditions. In this study, we investigated the mechanical behavior of monolithic foam glasses with similar total porosities but distinct mechanical responses. Our *in-situ* compression tests, combined with three-dimensional images from micro-CT scans, provided useful insights into the fracture behavior of the foams. During compression, porosity, mean equivalent diameter, and mean sphericity of the pores followed a standard behavior driven by densification, collapse, and deformation of internal structures. Fractures occurred under higher loads, particularly at the contact regions with the piston surfaces. This phenomenon was also evident in the strain maps obtained through DVC analysis, which described the microstructural changes and highlighted higher strain in these regions.

Higher porosity, larger pores and thicker pore walls were key factors in reducing the mechanical stability as assessed by Young's modulus and maximum compressive strength. We observed a power-law correlation between compressive strength and the percentage of large pores. Additionally, the uniformity of pore size distribution was identified as a potential factor influencing mechanical strength.

This work provides insights into the mechanical behavior and microstructural changes of foam glasses under various conditions. Our results suggest that optimizing pore size distribution, by minimizing large and irregularly sized pores, and ensuring enhanced structural integrity at external surfaces, can significantly improve the mechanical resistance of foam glasses while preserving their light weight and highly porous character. This knowledge can be leveraged to improve the applications of these materials by enhancing their stability and potentially preventing undesired damage.

## Acknowledgements

This study was partially funded by Coordenação de Aperfeiçoamento de Pessoal de Nível Superior - Brazil (CAPES) - Finance Code 001. Additionally, this project has received funding from the European Union's Horizon Europe research and innovation program under grant agreement no. 101131765 (EXCITE$^2$) for Transnational Access conducted at Institute of Mechanical Process Engineering and Mineral Processing (MVTAT) at TU Bergakademie Freiberg (TUBAF).

## AI use statement

OpenAI's ChatGPT (GPT-4-turbo) and Meta AI Llama 3.1 was used to improve grammar and readability of some sections of this manuscript. All scientific content, interpretations, and conclusions are the sole responsibility of the authors.